\documentclass[twocolumn, showpacs, english, preprintnumbers, amsmath, amssymb, superscriptaddress,aps]{revtex4-1}
\usepackage{latexsym}
\usepackage{epsfig,psfrag}
\usepackage{graphicx}
\usepackage{dcolumn}
\usepackage{bm}
\usepackage{color}
\usepackage{esint}
\usepackage{babel}
\usepackage{amsfonts}
\usepackage{enumerate}
\usepackage{mathbbol}
\definecolor{refkey}{RGB}{255,0,0}
\definecolor{labelkey}{RGB}{0,0,255}

\include{pdfsync}

\begin{document}

\title{Two-electron bound states near a Coulomb impurity in gapped graphene}

\author{Alessandro De Martino}
\affiliation{Department of Mathematics, City, University of London,
EC1V 0HB London, United Kingdom}
\author{Reinhold Egger}
\affiliation{Institut f\"ur Theoretische Physik,
Heinrich-Heine-Universit\"at, D-40225 D\"usseldorf, Germany}

\date{\today}

\begin{abstract}
We formulate and solve the perhaps simplest two-body bound state problem
for interacting Dirac fermions in two spatial dimensions. 
A two-body bound state is predicted for gapped graphene monolayers
in the presence of weakly repulsive electron-electron interactions and a 
Coulomb impurity with charge $Ze>0$, where the most interesting case corresponds to $Z=1$.
We introduce a variational Chandrasekhar-Dirac spinor wave function and 
show the existence of at least one bound state. This state leaves clear
signatures accessible by scanning tunneling microscopy.  One may
thereby obtain direct information about the strength of
electron-electron interactions in graphene.
\end{abstract}

\maketitle

\section{Introduction}
\label{sec1}

Ever since the Nobel prize winning experiments by the Manchester group in 2004 
\cite{geim2004,geim2005},  two-dimensional (2D)
graphene monolayers continue to attract a lot of attention.    
Besides graphene's application potential, much of the interest  comes from 
the fact that low-energy fermionic quasiparticles in graphene are 
governed by the 2D Dirac Hamiltonian \cite{rmp1,goerbig,philip,eva,kotov,miransky}. 
As a consequence, 
typical effects predicted by
relativistic quantum mechanics and/or quantum electrodynamics  
can be studied in table-top experiments.
Recent progress has demonstrated that one can reach the ballistic 
(disorder-free) regime \cite{Kim2016}, e.g., by 
encapsulating the graphene layer in boron nitride crystals \cite{boron}. 
We will therefore not consider disorder effects in this work.

Here we address the perhaps simplest interacting problem for relativistic fermions 
by considering a gapped graphene monolayer 
subject to relatively weak (screened) Coulomb interactions and 
in the presence of a single Coulomb impurity.
A band gap $2\Delta$ in the Dirac fermion spectrum may 
be caused by a variety of mechanisms, e.g., 
by strain engineering \cite{vozmediano}, spin-orbit coupling \cite{kanemele2005,huertas}, 
substrate-induced superlattices \cite{ponomarenko,song}, or simply as finite-size effect in 
a ribbon geometry \cite{rmp1}. 
The  Coulomb interaction strength is 
customarily quantified in terms of the effective fine-structure constant
\begin{equation}\label{alpha}
\alpha=\frac{e^2}{\kappa \hbar v_F} \simeq \frac{2.2}{\kappa},
\end{equation}
where $\kappa$ is a dielectric constant due to the surrounding substrate 
and $v_F\simeq 10^6$ m$/$s denotes the Fermi velocity \cite{kotov}.
The estimate in Eq.~\eqref{alpha} follows with $c/v_F\simeq 300$ and 
$e^2/(\hbar c)\simeq 1/137$.
The value of $\alpha$ thus depends on the choice of gating geometry and 
the ensuing screening effects.
As explained below, the weak interaction regime where our theory 
is applicable is defined by  $\alpha\alt 0.4$.
In this regime, we find that a pair of repulsively interacting 
Dirac fermions 
subject to the attractive potential of a Coulomb impurity with charge $Ze$ 
will form a two-body bound state localized near the impurity.  
We mainly focus on the most interesting case $Z=1$, but also comment
 on what happens for $Z>1$.  
 
The signatures of the predicted bound state could be 
 observed by means of scanning tunnelling microscopy (STM) experiments 
 similar to those previously reporting supercritical behavior in graphene \cite{crommie1,luican,crommie2}
 and trapped electron states in electrostatically defined graphene dots \cite{Lee2016,Gutierrez}.
An experimental observation of the predicted two-body bound state
could therefore probe and quantify Coulomb interaction effects.  
In order to consistently formulate this relativistic quantum-mechanical two-particle 
problem, it is necessary to stay away from the supercritical instability \cite{rmp1,kotov,novikov,gamayun},
otherwise one inevitably has to face a complicated many-body problem \cite{kotov}.
For small values of $\alpha$,  supercriticality is 
absent, and as explained below, a  two-particle bound state exists.
We mention in passing that the two-body problem in graphene has also been studied in 
Refs.~\cite{guinea,terekhov,koch,ziegler,entin,gaul,shytov,portnoi}.
In contrast to our work, however, those papers considered translationally invariant settings without Coulomb impurity.  Similar problems (again without Coulomb impurity) 
have also been analyzed in the context of 
Dirac excitons in single-layer transition-metal dichalcogenides \cite{rodin,zhong, trushin}.

The negatively charged two-electron hydrogen ion $H^-$ represents a 
classic problem of nonrelativistic quantum mechanics
 \cite{chandrasekhar,bethe,hill,bransden,andersen,sorba}.
In particular, it is well known \cite{hill} that $H^-$ has 
a single bound state in three spatial dimensions. 
The simplest way to prove its stability is to demonstrate the existence of a variational 
wave function with energy  below the ground-state energy of the hydrogen atom. 
Interestingly, it is impossible to achieve this task with a factorized wave function \cite{sorba}.
The simplest way to construct a variational wave function for the ground state of $H^-$ 
is due to Chandrasekhar \cite{chandrasekhar}. With $r_{l=1,2}=|{\bf r}_l|$ denoting
the distance of the respective electron from the proton, the Chandrasekhar ansatz for the 
spatial part of the two-particle wave function, $\Psi({\bf r}_1,{\bf r}_2)$,
contains two variational parameters ($a,b$) and is (up to a normalization factor) given by
\begin{equation}\label{chandr}
\Psi({\bf r}_1,{\bf r}_2)= e^{-a r_1 - b r_2} - \epsilon e^{ -b r_1-a r_2 }, 
\end{equation}
where $\epsilon=\mp 1$ corresponds to a spin singlet/triplet state, respectively. 
Here the important insight is that $a$ and $b$ are not required to be identical. 
Indeed, the minimal variational energy for a two-body bound state
is obtained for $a\neq b$ in the spin singlet configuration ($\epsilon=-1$). 
Improved variational energy estimates can be obtained by taking into account 
the dependence on the relative distance 
$r_{12}=|{\bf r}_1-{\bf r}_2|$ in Eq.~\eqref{chandr},
e.g., through a correlation factor of the form $(1+c\,r_{12})$ \cite{sorba}. 
Since the inclusion of such a factor is technically cumbersome yet not expected
to cause qualitative changes, cf.~Ref.~\cite{sorba},  we here restrict 
ourselves to uncorrelated wave functions and leave the analysis of correlation effects
to future work.

The nonrelativistic hydrogen ion has also been studied for the 2D case.
For instance, the so-called $D^-$ problem describes a donor impurity ion with 
two attached electrons in a 2D semiconductor quantum well 
\cite{phelps,louie,larsen92a,larsen92b,sandler,ivanov}. 
Interesting experimental
results on the $D^-$ problem have appeared in Ref.~\cite{etienne}, where
effects of quantum confinement on two-body bound-state energies have been studied.
In the absence of a magnetic field, only a single 
bound state in the spin singlet sector is expected again.
We note in passing that the $D^-$ problem is also similar 
to the negatively charged exciton ($X^-$) problem, which was
experimentally studied in quantum wells \cite{pepper}.

In this paper, we turn to the 2D relativistic counterpart of the above system, 
which is realizable in gapped graphene monolayers (or topological insulator surfaces) containing a Coulomb impurity. The corresponding relativistic $H^-$ problem is more difficult to define (and solve) because the single-particle Dirac Hamiltonian is 
unbounded from below \cite{brown,talman,nakatsuji}.  
It then appears at first sight as if two-particle  states of arbitrarily low energy 
can be generated by electron-electron interactions.  As discussed below, 
in  order to avoid this spurious and ultimately unphysical effect,  it 
 is necessary to employ a projection scheme which defines a mathematically 
clean framework.  Such a projection scheme can be devised for interacting
Dirac fermions in graphene if (i) a single-particle gap exists ($\Delta>0$),  
and (ii) electron-electron interactions are weak, 
see Refs.~\cite{sucher1,sucher2,haeusleregger,Egger}.

The structure of the remainder of this article is as follows.
 We introduce the Dirac-Coulomb model 
and the appropriate projection scheme in Sec.~\ref{sec2},
followed by a discussion of the variational approach to the two-body bound state
problem in Sec.~\ref{sec3}. (Details have been delegated to two appendices.)
To that end, we formulate a Chandrasekhar-Dirac spinor 
ansatz generalizing Eq.~\eqref{chandr} to the relativistic case. 
We evaluate all needed matrix elements  and discuss the variational 
bound-state energy  as a function of  $\alpha$.  In particular, we show 
variational estimates for the energy of the bound state in the presence of a standard
Dirac mass term causing a band gap, assuming a spin-singlet state and 
impurity charge $Z=1$.  We then turn to generalizations in Sec.~\ref{sec4}, 
where we shall address (i) what happens for impurity charge $Z>1$,  
(ii) the effects of a topologically nontrivial band gap as obtained, e.g., from spin-orbit coupling effects,  and (iii) the role of the valley state of each quasiparticle.  
In Sec.~\ref{sec5}, we address the observable consequences of the 
two-body bound state  accessible to STM experiments. 
Finally, we offer some conclusions in Sec.~\ref{sec6}.

\section{Dirac-Coulomb Model} 
\label{sec2}

We model the interacting two-particle problem for a gapped 
graphene monolayer in the presence of a charge impurity by the (properly projected, 
see below) Dirac-Coulomb Hamiltonian 
\begin{equation} \label{2bdiraccoulomb}
H  = \sum_{l=1,2} H_{\rm D}(l)  +V_{\rm 2b}.
\end{equation}
Here $H_{\rm D}(l)$ is the usual single-particle massive Dirac-Weyl 
Hamiltonian for particle $l=1,2$,
\begin{equation}\label{1bdiraccoulomb}
H_{\rm D}(l)  = H_0(l) + H_{\rm gap}(l) + V_{\rm 1b}(l), 
\end{equation}
with kinetic part (the index $l$ being understood) \cite{rmp1}
\begin{equation}\label{h0dirac}
H_0 = v_F ( \tau_z \sigma_x p_x + \sigma_y p_y ),
\end{equation}
where Pauli matrices $\bm \tau$ ($\bm \sigma$) act in valley 
(sublattice) space and the momentum operator is ${\bf p} =- i\hbar \nabla$.
With the dielectric constant $\kappa$, see~Eq.~\eqref{alpha}, the single-particle potential
{\begin{equation}
V_{\rm 1b} = -\frac{Ze^2}{\kappa r} = -\frac{Z\alpha \hbar v_F}{r} 
\end{equation}
describes a charge-$Z$ impurity at the origin.
We mainly focus on the most interesting case of unit charge, $Z=1$, 
but comment on the case $Z>1$ in Sec.~\ref{sec4a}. 
In $H_{\rm D}$, we also include the term 
\begin{equation} \label{diracmass}
H_{\rm gap} = \Delta \sigma_z ,
\end{equation}
which results in a topologically trivial band gap of size $2\Delta$ \cite{rmp1}.
However, it is straightforward to generalize our analysis to the case
of a spin- and valley-dependent topological band gap, e.g., due  
to an intrinsic spin-orbit coupling mechanism \cite{kanemele2005,huertas},
\begin{equation}
  H_{\rm so} = \Delta \sigma_z \tau_z s_z,
\label{somass}
\end{equation}
where $\bm s$ are Pauli matrices in spin space, see Sec.~\ref{sec4b}. Note that
$H_{\rm gap}$ has the same sign for both spin projections while 
$H_{\rm so}$ has opposite sign. 

The operator $H_{\rm D}$ is Hermitian only for $Z<Z_{\rm crit}=1/(2\alpha)$.
Indeed, the lowest single-particle bound-state energy becomes imaginary  
for $Z>Z_{\rm crit}$, see Eq.~\eqref{lowesteigenvalue} in App.~\ref{appA}.
 We note that by regularizing the $r\to 0$  singularity of the Coulomb
 potential, the threshold shifts to a larger value $\tilde Z_{\rm crit}$, 
which weakly depends on the precise regularization prescription \cite{gamayun}.
Even in the regularized scheme, however, Hermiticity is lost for $Z>\tilde Z_{\rm crit}$.
In the latter regime, bound states of the regularized Hamiltonian $\tilde H_{\rm D}$ 
dive into the negative part 
of the continuum spectrum ($E<-\Delta$) one by one, and thereby become quasi-stationary states, see  Ref.~\cite{gamayun}.  As remarked in Sec.~\ref{sec1}, 
one then necessarily has to consider the full many-body problem for $Z>\tilde Z_{\rm crit}$. For this reason, we will restrict ourselves to the weak-coupling regime. 
Moreover, for the sake of transparency, we focus on the narrower range
 $Z<1/(2\alpha)$, where no need for regularization arises and the exact 
Dirac-Coulomb wave functions summarized in App.~\ref{appA} can be used.
For the most interesting case $Z=1$, this implies that our theory 
is at best applicable for $\alpha<1/2$. 

Since $H_{\rm D}$ is diagonal in valley and spin space,
we can always choose a factorized form for the single-particle 
wave functions,
\begin{equation}
\Psi_{\tau,s}(x,y) =\Psi(x,y) | \tau , s \rangle,
\end{equation}
where $\Psi(x,y)$ refers to the spatial part (including sublattice space), and 
$ | \tau , s \rangle$ with $\tau = K/K'=\pm 1$ and $s=\uparrow/\downarrow =\pm 1$ 
denotes the eigenstates of the operator $\tau_z s_z$.
For an eigenstate $\Psi_{\tau,s}$
of $H_{\rm D}$ with energy $E$, the  symmetry relations 
\begin{equation}
 \sigma_y \Psi_{\tau,s}^* =\Psi_{\tau,-s},\quad
\sigma_y \Psi_{\tau,s}  =\Psi_{-\tau,s}, \quad
\Psi_{\tau,s}^* =\Psi_{-\tau,-s},
\end{equation} 
 are a manifestation of the well-known fourfold spin-valley degeneracy of each energy level \cite{rmp1}. 

Next, the two-body Coulomb interaction in Eq.~\eqref{2bdiraccoulomb} is formally given by (see App.~\ref{appB})
\begin{equation}
V_{\rm 2b} =\frac{e^2}{\kappa|{\bf r}_{1}-{\bf r}_2|}.
\end{equation}
However, since the spectrum of the single-particle Dirac Hamiltonian is unbounded 
from below, the many-body Dirac-Coulomb 
Hamiltonian \eqref{2bdiraccoulomb} cannot have true two-particle bound states.
This problem was pointed out long ago by Brown and Ravenhall in their study of 
relativistic effects in the helium atom \cite{brown}. 

A rigorous procedure to construct a mathematically well-defined many-particle Hamiltonian  for interacting massive Dirac fermions in three spatial dimensions was
devised by Sucher \cite{sucher1,sucher2}. Fortunately, as has been shown in 
detail in Refs.~\cite{haeusleregger,Egger}, Sucher's approach can be readily adapted 
to the case of 2D Dirac fermions in graphene as long as the single-particle
spectrum exhibits a band gap. In effect, $H$ in Eq.~\eqref{2bdiraccoulomb} thereby has 
to replaced by the projected Hamiltonian \cite{haeusleregger,Egger}
\begin{equation}\label{Hproj}
 H_+= H_{\rm D}(1) + H_{\rm D}(2) + \Lambda_+ V_{\rm 2b} \Lambda_+,
\end{equation}
with the projection operator $\Lambda_+=\Lambda_+(1)\Lambda_+(2)$. 
Here, using ${\cal E}(l)=[H_{\rm D}^2(l)]^{1/2}$, the single-particle operator
 $\Lambda_+(l) = \left[{\cal E}(l)+H_{\rm D}(l)\right]/2{\cal E}(l)$
projects onto the space spanned by the positive-energy eigenstates of $H_{\rm D}(l)$.
As detailed in Refs.~\cite{sucher1,sucher2,haeusleregger,Egger}, the projected Hamiltonian $H_+$ takes into account the most 
important effects of the electron-electron interaction.  In fact, due to the presence
of a band gap, the replacement $H\to H_+$ does not introduce approximations
concerning the ground state of the system in the limit of weak Coulomb repulsion.   Moreover, the projection guarantees that the Hamiltonian $H_+$ 
can possess bona-fide two-particle bound states.  
 For instance, in the nonrelativistic limit realized for
energies very close to the upper band edge, 
by expanding Eq.~\eqref{Hproj} to lowest nontrivial order in $1/\Delta$,  
we obtain  (up to a constant energy 
shift $2\Delta$) the Schr\"odinger Hamiltonian
\begin{equation}\label{schrodinger}
H_{\rm S}= \sum_{l=1,2} \left( \frac{{\bf p}^2_l}{2m} + V_{\rm 
1b}(l) \right) + V_{\rm 2b},
\end{equation}
with the mass $m=\Delta/v_F^2$. Equation \eqref{schrodinger}
describes a $D^-$ center in a 2D semiconductor quantum well and 
has been studied in Refs.~\cite{phelps,louie,larsen92a,larsen92b,sandler,ivanov}.
One can therefore regard $H_+$ in Eq.~\eqref{Hproj} 
as a natural relativistic generalization 
of the $D^-$ impurity center problem.

In the remainder of this paper, we shall employ units with $\hbar =v_F=1$.

\section{Variational approach}
\label{sec3}

The Hamiltonian $H_+$ acts in the tensor space of two copies of the 
eight-dimensional single-particle Hilbert space. 
For the non-interacting system with $V_{\rm 2b}=0$,
two-particle spinor wave functions for bound states are written as
antisymmetrized products of single-particle wave functions,
\begin{equation}
\Phi({\bf r}_1,{\bf r}_2) = {\cal A} \left[ 
\Psi^{(1)}_{\tau_1,s_1}({\bf r}_1) \otimes \Psi^{(2)}_{\tau_2,s_2}({\bf r}_2) \right],
\end{equation}
where ${\cal A}$ is the antisymmetrization operator and $\Psi^{(l)}_{\tau_l,s_l}$ is a single-particle eigenstate with eigenenergy $E_{n_l,j_l}$ for the 2D relativistic hydrogen problem described by $H_{\rm D}(l)$. 
To keep the paper self-contained, we summarize the well-known solution of $H_{\rm D}$ in App.~\ref{appA}. Eigenstates are labelled
by the principal quantum number $n=0,1,2,\ldots,$ and by the half-integer angular
momentum $j$. The ground state of $H_{\rm D}(l)$ 
is realized for $n_l=0$ and $j_l=1/2$. Hence 
the noninteracting ($V_{\rm 2b}=0$) two-particle ground state has the energy 
$E_{\rm gs} = 2 E_{0,\frac{1}{2}}$, where both particles occupy
the respective single-particle ground state. This two-particle 
state has finite total angular momentum $j=1$, where the angular 
momentum operator is $J_z=J_z(1)+J_z(2)$ with
 $J_z(l)=-i\partial_{\theta_l}+\sigma_z(l)/2$. 
 
To study the ground state of the interacting system, one could attempt to
treat the two-body Coulomb repulsion by perturbation theory. However, 
one then finds that for $Z=1$, the resulting binding energy is always negative. 
In other words, first-order
perturbation theory in the Coulomb interaction incorrectly predicts that there is no bound state (see below),
and one has to proceed in a nonperturbative manner to investigate this issue.
We here employ a variational treatment and construct a relativistic version 
of the  Chandrasekhar wave function \eqref{chandr},
dubbed Chandrasekhar-Dirac ansatz.
We shall see that the corresponding energy functional is bounded from below
and thus provides a variational estimate of the binding energy.

In this section, we focus on a valley- and spin-independent band gap, see 
Eq.~(\ref{diracmass}). The topological band gap term in Eq.~\eqref{somass}
then only requires a few adjustments, see Sec.~\ref{sec4b}.
Moreover,  we assume here that both quasiparticles are in the
same valley but show in Sec.~\ref{sec4c} that the variational result does
not change for quasiparticles in opposite valley states.
Since the Hamiltonian commutes with both $S_z$ and ${\bf S}^2$, 
where ${\bf S}={\bf s}(1)+{\bf s}(2)$ is (twice) the total spin operator, 
we have a spin singlet and a spin triplet state, where in
 the first (second) case, the spatial part of the wave function must be 
 symmetric (antisymmetric). We shall see that for $Z=1$, the Chandrasekhar-Dirac ansatz 
 predicts a bound state in the singlet but not in the triplet channel. 
 However, for $Z>1$, bound states are found in both cases.

\subsection{Chandrasekhar-Dirac ansatz}

Our Chandrasekhar-Dirac ansatz is formulated as follows.
We assume that the two-particle wave function has a factorized form,
$\Phi_{\rm tot} =\Phi |\chi \rangle,$
where $|\chi \rangle$ is the normalized spin part (singlet or triplet) and 
$\Phi$ the spatial part,
\begin{equation}\label{cdans}
\Phi({\bf r}_1, {\bf r}_2) = \Psi_{I}({\bf r}_1) \Psi_{O}({\bf r}_2) - \epsilon 
\Psi_{O}({\bf r}_1) \Psi_{I}({\bf r}_2) ,
\end{equation}
where $\Psi_I$ and $\Psi_O$ are the normalized ground-state
eigenspinors of the 2D relativistic hydrogen problem, see 
App.~\ref{appA}, with $Z$ replaced by  
variational parameters $Z_I$ and  $Z_O$, respectively.  
The parameter $\epsilon=\mp 1$ corresponds to the spin singlet/triplet sector, and
the valley part of the wave function is understood. 
We mention in passing that in the triplet case, a wave function composed of two ground-state 
single-particle orbitals might not represent the optimal choice, 
see Sec.~\ref{sec4a}.

Since the single-particle Hamiltonian does not depend on the spin projection, 
we can use the same wave function for both particles. Explicitly, the spinors with $\lambda=I,O$ have the  form (see App.~\ref{appA})
\begin{equation}\label{eigenspin}
\Psi_\lambda({\bf r}) ={\cal N}_\lambda
r^{\gamma_\lambda-1/2} e^{-p_\lambda r}
\left(\begin{array}{l}
1\\ i\kappa_\lambda e^{i\theta}
\end{array}\right),
\end{equation}
where
\begin{eqnarray} \nonumber
\gamma_\lambda &=& \sqrt{\frac{1}{4}-Z^2_\lambda\alpha^2},\qquad 
p_\lambda = 2 \Delta Z_\lambda\alpha, \\
\kappa_\lambda &=& \sqrt{\frac{1-2\gamma_\lambda}{1+2\gamma_\lambda}} = 
\frac{Z_\lambda\alpha}{\frac{1}{2}+\gamma_\lambda}= \frac{\frac{1}{2}-\gamma_\lambda}{Z_\lambda\alpha} .
\end{eqnarray}
The normalization constant is given by
\begin{equation}
{\cal N}_\lambda = \sqrt{ \frac{(2p_\lambda)^{2\gamma_\lambda+1}}{2\pi (1+\kappa_\lambda^2)\Gamma(2\gamma_\lambda+1)} }
\end{equation}
with the Gamma function $\Gamma(x)$.
Equation \eqref{eigenspin} represents the ground-state eigenspinor of the single-particle
Dirac Hamiltonian
\begin{equation}
H_\lambda = -i{\bm \sigma}\cdot \nabla +\Delta \sigma_z -\frac{Z_\lambda\alpha}{r},
\end{equation}
with eigenvalue $E=2\Delta\gamma_\lambda =\Delta \sqrt{1-4Z_\lambda^2 \alpha^2}.$
Note that the spinors $\Psi_\lambda$ are not orthogonal. Their overlap 
$S = \langle \Psi_I |\Psi_O \rangle $ is given by
\begin{eqnarray}
S &=&\frac{(1+\kappa_I\kappa_O)}{\sqrt{(1+\kappa^2_I)(1+\kappa^2_O)}}
\frac{\Gamma(\gamma_I+\gamma_O+1)}{\sqrt{\Gamma(2\gamma_I+1)\Gamma(2\gamma_O+1)}} \nonumber \\
&\times& \frac{(2p_I)^{\gamma_I+1/2}(2p_O)^{\gamma_O+1/2}}{(p_I+p_O)^{\gamma_I+\gamma_O+1}}.
\label{overlap}
\end{eqnarray}

\subsection{Energy functional}

We now evaluate the energy functional 
\begin{equation}\label{Efunctional}
E_{\epsilon}(Z_I,Z_O)=\frac{\langle \Phi_{\rm  tot} | H_+ | \Phi_{\rm tot} \rangle }{\langle \Phi_{\rm tot} | \Phi_{\rm tot} \rangle}
=\frac{\langle \Phi | H_+ | \Phi \rangle }{\langle \Phi | \Phi \rangle},
\end{equation}
with $H_+$ in Eq.~(\ref{Hproj}).
The index $\epsilon=\mp$ in $E_\epsilon$ refers to the spin singlet/triplet state, and the normalization factor is given by
\begin{eqnarray}
\langle \Phi_{\rm tot} | \Phi_{\rm tot} \rangle &=& 
\langle \Phi | \Phi \rangle  = 2 \langle \Psi_I |\Psi_I \rangle
\langle \Psi_O |\Psi_O \rangle - 2 \epsilon\langle \Psi_I |\Psi_O \rangle^2 
\nonumber \\
& =& 2\left(1 -\epsilon S^2\right).
\end{eqnarray}
Next, the matrix element of the single-particle Hamiltonian has the form
\begin{eqnarray}
&&\sum_{l=1,2}\langle \Phi_{\rm tot} |  H_{\rm D}(l) | \Phi_{\rm tot} \rangle  = \\
\nonumber
&&=2\left(\sum_\lambda \langle \Psi_\lambda | H_{\rm D} |  \Psi_\lambda \rangle -
2 \epsilon \langle \Psi_I | H_{\rm D} |  \Psi_O \rangle S \right),
\end{eqnarray}
where we have used the normalization of the one-particle spinors. 
By writing the single-particle Hamiltonian as
\begin{equation}
H_{\rm D} =H_\lambda+ (Z_\lambda-Z)\frac{\alpha}{r} ,
\end{equation}
one can directly evaluate the matrix elements. We obtain
\begin{eqnarray}
\langle \Psi_\lambda | H_{\rm D} | \Psi_\lambda \rangle &=& 
 2\Delta \gamma_\lambda  + (Z_\lambda-Z) {\cal V}_\lambda, \\ \nonumber
\langle \Psi_{\bar \lambda} | H_{\rm D}| \Psi_\lambda \rangle &
=& 2\Delta \gamma_\lambda S + 
(Z_\lambda-Z) {\cal U}  \\  \nonumber
&=&2\Delta \gamma_{\bar \lambda} S + 
(Z_{\bar \lambda}-Z) {\cal U},
\end{eqnarray}
where $\bar \lambda=O$ for $\lambda=I$ and vice versa, and
\begin{eqnarray}
{\cal V}_\lambda &=& \langle \Psi_\lambda |(\alpha/r_1)| 
\Psi_\lambda \rangle  =\alpha\frac{p_\lambda}{\gamma_\lambda} , \\ \nonumber
{\cal U} &=& \langle \Psi_I |(\alpha/r_1)| \Psi_O \rangle  =
\alpha \frac{p_I+p_O}{\gamma_I+\gamma_O}S .
\end{eqnarray}
It is reassuring to note that the minimum with respect to $Z_\lambda$ for the energy
\begin{equation}
 \langle \Psi_\lambda | H_{\rm D} | \Psi_\lambda \rangle 
 = 2\Delta \sqrt{1/4-Z_\lambda^2\alpha^2} + 
\frac{2\Delta  Z_\lambda (Z_\lambda-Z) \alpha^2}{\sqrt{1/4-Z^2_\lambda\alpha^2}}
\end{equation}
occurs exactly at $Z_\lambda=Z$ provided that $Z<Z_{\rm crit}=1/(2\alpha)$.
Since we have used the exact structure of the Dirac-Coulomb wave function, 
the result reproduces the exact ground-state energy, $E_{\rm gs}=2\Delta \gamma$. 

We now proceed with the two-body matrix element, see also App.~\ref{appB},
\begin{eqnarray}
{\cal V}_{\rm 2b} &=& \langle \Phi_{\rm tot} | (\alpha/r_{12}) | \Phi_{\rm tot} \rangle =
 2 \left( {\cal V}_{\rm 2b}^{\rm dir} -
\epsilon {\cal V}^{\rm exc}_{\rm 2b} \right),\\
 \nonumber {\cal V}_{\rm 2b}^{\rm dir}  &=& 
 \int d{\bf r}_1d{\bf r}_2 \, \left| \Psi_I({\bf r}_1)\right|^2 \frac{\alpha}{r_{12}}
 \left| \Psi_O({\bf r}_2)\right|^2 , \\ \nonumber
 {\cal V}^{\rm exc}_{\rm 2b} & = & 
\int d{\bf r}_1 d{\bf r}_2 \left[   \Psi^\dagger_I({\bf r}_1)  
\Psi_O({\bf r}_1)  \right]  \frac{\alpha}{r_{12}}  \left[   \Psi^\dagger_O({\bf r}_2)  
\Psi_I({\bf r}_2)  \right].
\end{eqnarray}
The standard procedure to evaluate the multiple integrals in 
${\cal V}^{\rm dir}_{\rm 2b}$ and ${\cal V}^{\rm exc}_{\rm 2b}$ is to use a 2D 
partial wave expansion of $1/r_{12}$. However, the resulting series representation 
converges only very slowly. Following Ref.~\cite{sandler},
we found it more convenient to use an integral representation in terms 
of elliptic functions,
\begin{widetext}
\begin{eqnarray}
{\cal V}_{\rm 2b}^{\rm dir}  & =& \frac{2\alpha}{\pi} 
 \frac{p_I^{2\gamma_I+1}p_O^{2\gamma_O+1}}{
 (\gamma_I +\gamma_O+\frac{1}{2})
 B(2\gamma_I+1,2\gamma_O+1)} \int_0^1 ds \, {\bf K}(s)\left[ 
 \frac{s^{2\gamma_O}}{(p_I+sp_O)^{2(\gamma_I+\gamma_O)+1}} + 
 \frac{s^{2\gamma_I}}{(sp_I+p_O)^{2(\gamma_I+\gamma_O)+1}} 
 \right] , \label{directTerm} \\
{\cal V}^{\rm exc}_{\rm 2b} 
&=& \frac{8\alpha}{\pi} \frac{(p_I+p_O)S^2}{(\gamma_I+\gamma_O) 
B(\gamma_I+\gamma_O,\gamma_I+\gamma_O)}
 \int_0^1 ds \, {\bf K}(s)
 \frac{s^{\gamma_I+\gamma_O}}{(1+s)^{2(\gamma_I+\gamma_O)+1}},
 \label{exchangeTerm}
\end{eqnarray}
\end{widetext}
where $B(x,y)=\Gamma(x)\Gamma(y)/\Gamma(x+y)$ is Euler's beta function 
and ${\bf K}(s)$ denotes the complete elliptic integral of first kind \cite{olver}. 

Importantly, we here calculated the matrix elements of the full interaction 
operator rather than those of the projected operator,
$\Lambda_+ (\alpha/r_{12}) \Lambda_+$,
which are more difficult to obtain and would require a 
detailed numerical analysis.
Both matrix elements coincide if the trial wave 
function has vanishing projection onto
the negative energy eigenfunctions of $H_{\rm D}$. We show in Sec.~\ref{validity}
that this is in general not the case, and hence using the unprojected Coulomb 
interaction is strictly speaking not justified. However, the energy functional 
turns out to be bounded from below and does predict a two-particle bound state. 
More importantly, we have verified that for $\alpha\alt 0.4$ the cumulative 
weight of negative energy states in our trial wave 
function is very small  ($\alt 1\%$), see Sec.~\ref{validity}. 
Indeed, negative energy states will only be important if 
typical interaction matrix elements can overcome the band gap $2\Delta$.
For small $\alpha$, one therefore expects at most small quantitative corrections
in the bound-state energy because of this approximation.   
For a treatment of stronger interactions with $0.4\alt\alpha<1/2$, 
 however, one must resort to the matrix elements of the projected two-body operator. 
We leave this task for future work.

We now collect all terms and obtain 
\begin{eqnarray}
E_{\epsilon}(Z_I,Z_O) &=& 
\sum_\lambda \left( 2\Delta \gamma_\lambda +
\frac{(Z_\lambda-Z)({\cal V}_\lambda-\epsilon S {\cal U})}{1-\epsilon S^2}\right) \nonumber\\
&+&\frac{{\cal V}^{\rm dir}_{\rm 2b} - \epsilon {\cal V}^{\rm exc}_{\rm 2b}}{1-\epsilon S^2}.
\label{energyfunct}
\end{eqnarray}
The energy functional $E_\epsilon(Z_I,Z_O)$ has the following interesting features. 
First of all, $E_\epsilon(Z_I,Z_O)$ is symmetric under an exchange of
its arguments. 
Second, as illustrated in Fig.~\ref{enfunctS} for the spin-singlet case,
this energy is bounded from below as long as $Z\alpha< 1/2$.
Third, for small $\alpha$, we have checked that
$E_{\epsilon}(Z_I,Z_O)$ reduces to the corresponding nonrelativistic energy functional 
for the $D^-$ problem in 2D semiconductors \cite{sandler}. 
However, in contrast to the nonrelativistic case, $E_{\epsilon}(Z_I,Z_O)$ is not 
homogeneous in $\alpha$, and hence the bound-state energy explicitly 
depends on $\alpha$.  
As in the nonrelativistic case, this energy 
 minimum is realized for unequal values of $Z_I$ and $Z_O$.

With $\gamma=\sqrt{1/4-Z^2\alpha^2}$, the binding energy of the two-body bound state 
is defined for the optimized choice of $Z_{I,O}$ as
\begin{eqnarray}\nonumber
E_{\epsilon,b}(Z_I,Z_O) &=& \Delta(1+2\gamma) - E_\epsilon(Z_I,Z_O) \\ 
&\equiv& \frac{ \alpha^2\Delta}{2} \bar E_{\epsilon,{\rm b}}(Z_I,Z_O) ,
\label{ebinding}
\end{eqnarray}
where $\Delta(1+2\gamma)$ denotes
the energy of a state in which one of the particles is in the  
ground state of $H_{\rm D}$ and the other in the lowest 
positive energy state of the continuum spectrum of $H_{\rm D}$,
just above the gap. Equation \eqref{ebinding} defines
the rescaled dimensionless binding energy $\bar E_{\epsilon,{\rm b}}$ 
(in units of $\alpha^2\Delta/2$).  
In the singlet case, $\bar E_{\rm -,b}$  approaches 
the nonrelativistic value $\bar E_{\rm -,b}^{(0)}=0.307$ \cite{sandler}
for $\alpha\to 0$.  For finite $\alpha$, deviations of $\bar E_{\rm -,b}$ from 
$\bar E_{\rm -,b}^{(0)}$ indicate the importance of relativistic effects.  

\begin{figure}[t]
\centering
\includegraphics[width=.5\textwidth]{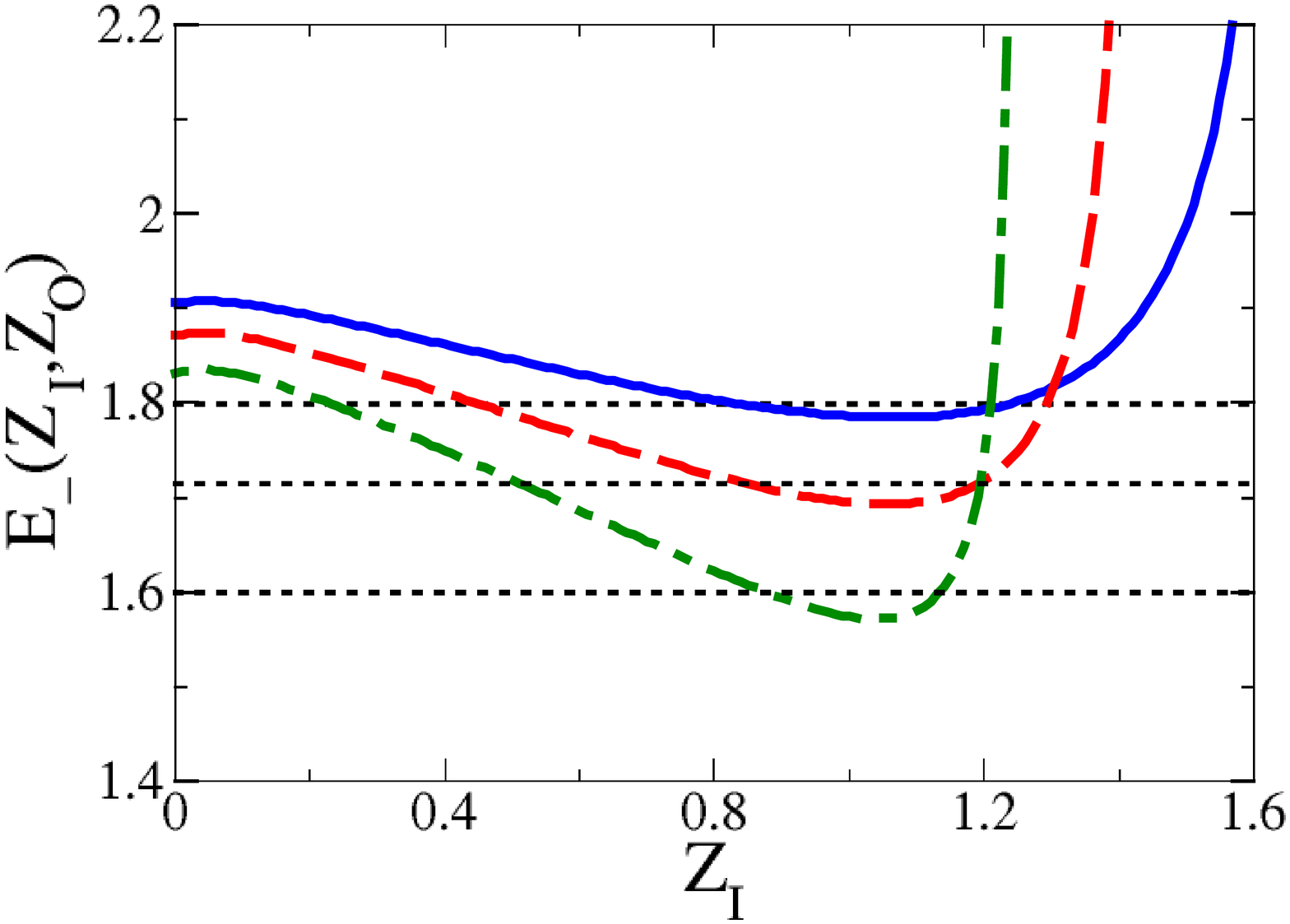}
\caption{ 
Energy functional $E_{\epsilon=-1}(Z_I,Z_O)$ in units of $\Delta$ 
for the spin singlet state as a function of $Z_I$ for $Z=1$, taking $Z_O=0.3$. 
The green dash-dotted (red dashed, blue solid) curve is for 
 $\alpha=0.3~(0.35,~0.4)$, respectively. 
Horizontal lines indicate the respective threshold energies $\Delta(1+2\gamma)$. 
When a minimum exists below the threshold (as is the case for all shown $\alpha$), 
a bound state is present.} 
\label{enfunctS}
\end{figure}

Figure \ref{enfunctS} shows that the energy functional 
for the singlet state with $Z=1$ has a minimum for all studied values of $\alpha$.
Moreover, the energy minimum is located below the threshold, i.e.,
the binding energy is positive, and we have a two-body bound state in the spin singlet sector. 
In contrast to that, our variational approach predicts that the energy 
functional has a minimum also for the spin triplet sector, but the minimum is now above the 
threshold and thus does not describe a bound state. 
We also notice that the energy functional for $Z_I=Z_O=Z$ simply yields 
the ground-state energy of a two-particle state with the Coulomb repulsion treated 
within first-order perturbation theory. As anticipated above, we find $E_{-}(1,1)>\Delta(1+2\gamma)$, 
and therefore perturbation theory is not able to correctly 
describe the bound state for $Z=1$.

\begin{table}
 \begin{tabular}{|c|c|c|c|}  \hline
  $\alpha$ &  $\bar E_{\rm -,b}$ & $Z_O$ & $Z_I$  \\   \hline\hline
  $0.01$	 & $0.307$ & $0.289$ & $1.090$	  \\    \hline
  $0.05$	& $0.308$ & $0.290$ & $1.090$ \\  \hline
  $0.10$	&  $0.310$ & $0.292$ & $1.088$ \\\hline
  $0.15$	& $0.314$ & $0.295$ & $1.086$ \\     \hline
  $0.20$	& $0.320$ & $0.299$ & $1.081$  \\   \hline
  $0.25$	& $0.327$ & $0.305$ & $1.076$ \\  \hline
  $0.30$	& $0.337$ & $0.314$ & $1.068$ \\  \hline
  $0.35$	& $0.350$ & $0.325$ & $1.058$ \\   \hline
  $0.40$	& $0.366$ & $0.341$ & $1.045$  \\     \hline  
    \end{tabular}
\caption{Rescaled binding energy $\bar E_{\rm -,b}$, see Eq.~\eqref{ebinding}, 
 for the two-body bound state in the spin singlet sector with $Z=1$  and 
several values of $\alpha$.  
These values are plotted in Fig.~\ref{figure2}. We also specify the 
effective charges $Z_O$ and $Z_I$ minimizing the
 energy functional, cf.~Fig.~\ref{figure3}. } 
\label{table}
\end{table}

Table \ref{table} lists the binding energy in the spin singlet 
sector for several values of $\alpha$. 
Note that both $Z_I$ and $Z_O$ are (well) below the critical value 
$Z_{\rm crit}=1/(2\alpha)$ for all cases considered in Table \ref{table}.
The fact that for $Z=1$ the minimum happens to be at $Z_O<1$ and $Z_I>1$ 
(or vice versa, due to the symmetry of $E_\epsilon$) is rationalized by noting that 
only in this case the two factors $(Z_I-1){\cal V}_I$ and $(Z_O-1){\cal V}_O$ in 
Eq.~\eqref{energyfunct} will have opposite sign. 
Physically, one quasiparticle then partially screens the impurity charge seen by the 
other quasiparticle.

We observe from Table~\ref{table} that relativistic effects tend to increase 
the binding energy.  Approximately, we find the scaling
 $\bar E_{\rm -,b}-\bar E_{\rm -,b}^{(0)}\sim\alpha^2$ illustrated in Fig.~\ref{figure2}.
It is interesting to note that the variational parameter $Z_O$ increases with 
$\alpha$ while $Z_I$ decreases,  see Fig.~\ref{figure3}. 
Since the atomic Bohr radius is  $\sim 1/Z\alpha$, we conclude that in the 
relativistic case the outer (inner) electron will be closer to (further away from)
the nucleus than in the nonrelativistic case.   

\begin{figure}[t]
\includegraphics[width=0.5\textwidth]{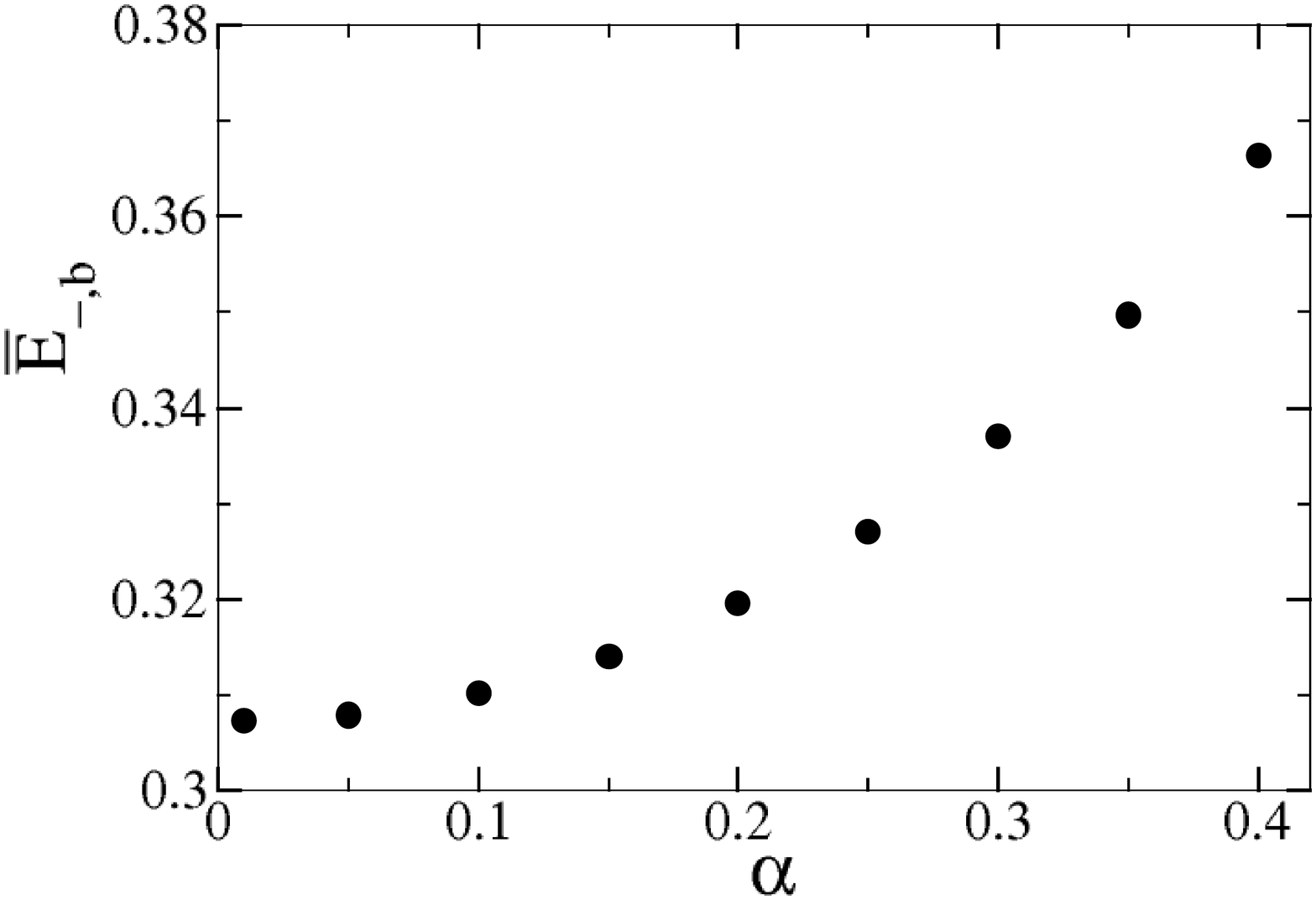}
\caption{Rescaled dimensionless binding energy $\bar E_{\rm -,b}$ vs $\alpha$, 
see Eq.~\eqref{ebinding}, in the spin singlet sector with $Z=1$.} 
\label{figure2}
\end{figure}

\begin{figure}[t]
\includegraphics[width=.5\textwidth]{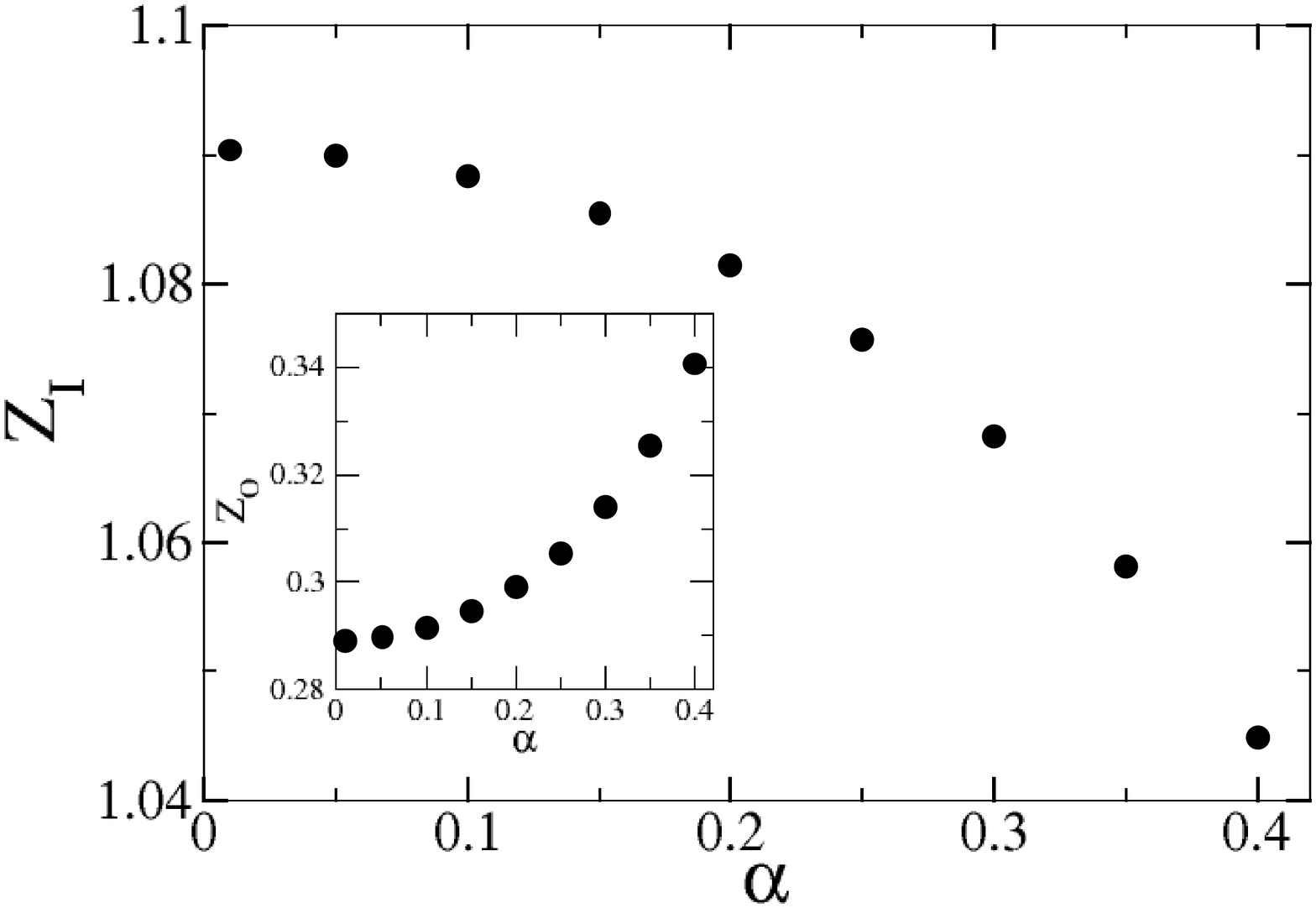}
\caption{Optimal values of $Z_I$ (main panel) and $Z_O$ (inset) 
vs $\alpha$ for the spin singlet case with $Z=1$.}
\label{figure3}
\end{figure}

\subsection{Validity of the variational approach}
\label{validity}

Before we conclude this section, we comment on the validity of this 
variational calculation. First, we observe that the 
single-particle orbital $|\Psi_\lambda\rangle$ is the normalized 
ground state of  a modified Dirac Hamiltonian $H_\lambda$ which is 
related to $H_{\rm D}$ by
 \begin{equation}
H_{\rm D} = H_\lambda +  \frac{(Z_\lambda-Z)\alpha}{r}.
\end{equation}
Assuming that $|\Psi_{E,j} \rangle$ is in 
the continuous spectrum of $H_{\rm D}$, we have
\begin{equation}
(E-E_\lambda)\langle \Psi_\lambda | \Psi_{E,j} \rangle =  
\langle \Psi_\lambda |[(Z_\lambda-Z)\alpha/r]  |\Psi_{E,j} \rangle.
\end{equation}
As a consequence, for $E_\lambda\ne E$
and $Z_\lambda=Z$,  the state 
$|\Psi_\lambda \rangle$ is orthogonal to $|\Psi_E\rangle$.
This is to be expected since in that case, they are eigenstates
of the same Hermitian operator with different eigenvalues.
However, for $Z_\lambda \neq Z$, both states will generically have a finite overlap.
Since we take $|\Psi_\lambda\rangle$ as the ground-state orbital, it is clear that the 
overlap with states in the negative-energy continuum will be
suppressed by a factor $\approx 1/\Delta$. 
Actually, the overlap $\langle \Psi_\lambda  |\Psi_{E,j} \rangle$ can be evaluated analytically, 
see App.~\ref{appA}. We can thereby compute
the total weight of our variational wave function 
on the negative energy states,
\begin{equation}\label{wweight}
W_\lambda(\alpha)=\int_{E<-\Delta} \frac{dE}{2\pi\sqrt{E^2-\Delta^2}} \left| \langle \Psi_\lambda  |\Psi_{E,j} \rangle  \right|^2.
\end{equation}
The result is shown in Fig.~\ref{weight} for several values of 
$\alpha$ and the corresponding
$Z_\lambda$ from Table~\ref{table}. We find that the 
total weight $W_\lambda(\alpha)$ for $\alpha \leq 0.4$ is 
at most of order $0.01$.  It then stands to reason that neglecting the 
projection operator in the evaluation of Coulomb matrix elements does not 
significantly affect the variational estimate of 
the binding energy for $\alpha\alt 0.4$.

\begin{figure}[t]
\includegraphics[width=0.5\textwidth]{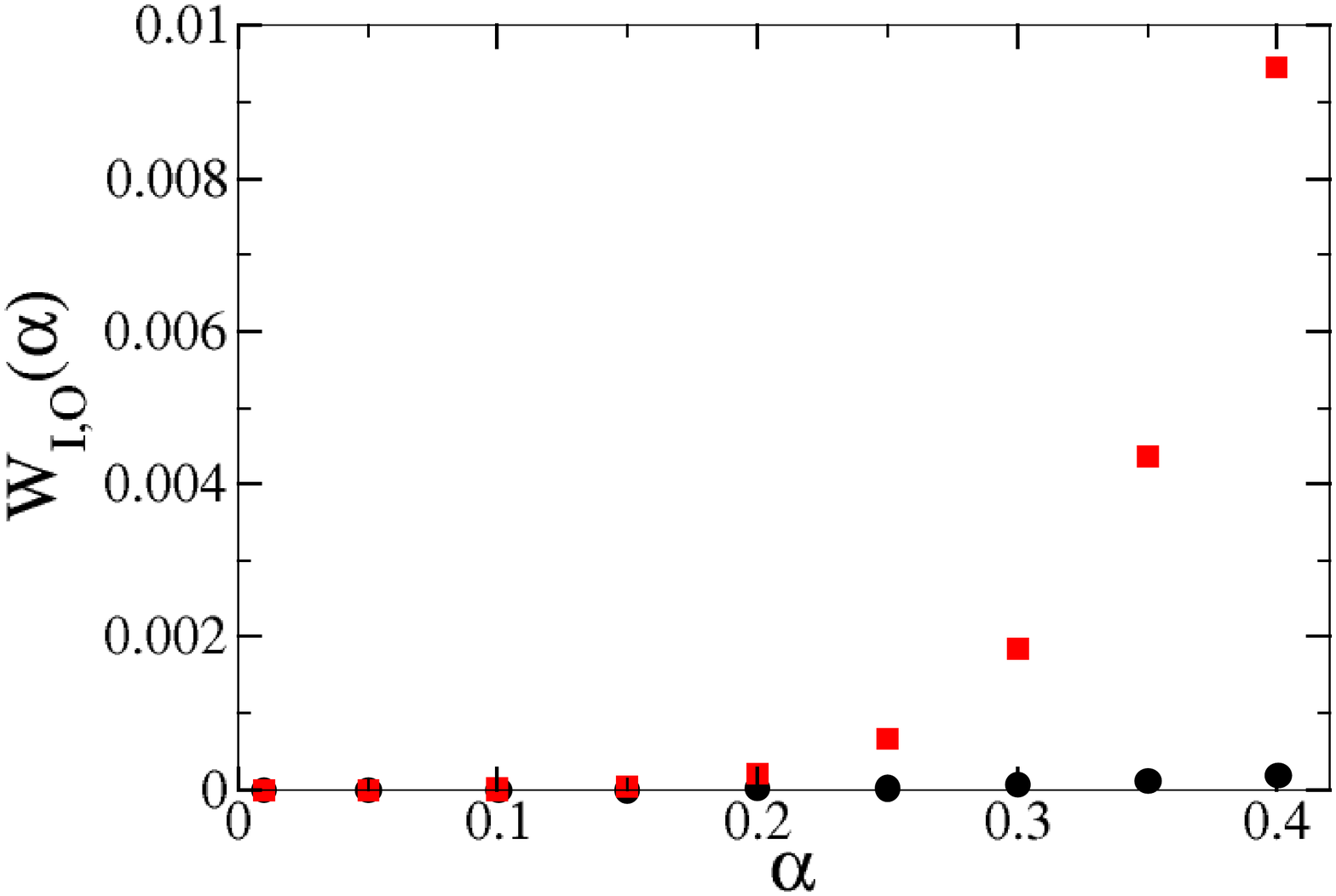}
\caption{Cumulative weight $W_\lambda$ vs $\alpha$, see Eq.~\eqref{wweight}, 
of the negative energy states in the variational wave function for $Z=1$,
where $\lambda=I$ ($\lambda=O$) corresponds to black circles (red squares).}
\label{weight}
\end{figure}

\section{Generalizations}
\label{sec4}

So far we have restricted ourselves to the case of two quasiparticles 
in the same valley, in the presence of a band gap, and for 
an impurity of charge $Z=1$. In this section, we briefly
address various extensions, namely (i) the case of an impurity with charge $Z>1$,  
(ii) a topological band gap, and (iii)  quasiparticles in different valleys.

\subsection{Impurity charge $Z>1$}
  \label{sec4a}

From a theoretical perspective, the case $Z>1$ is less interesting than $Z=1$ 
because a bound state is then found already in perturbation theory.
In fact, when treating the two-body Coulomb interaction perturbatively,
the energy of the lowest singlet state, taken in the simple factorized form
$\Phi_{\rm tot}= \Psi_{0,\frac{1}{2}}({\bf r}_1)  \Psi_{0,\frac{1}{2}}({\bf r}_2)| \chi \rangle$, coincides with the value of the energy functional $E_\epsilon(Z_I,Z_O)$ 
for $\epsilon=0$ and $Z_I=Z_O=Z$,
\begin{equation}\label{pertE}
E_{\rm pert}= E_{0}(Z,Z)=4\Delta \gamma + {\cal V}_{\rm 2b}^{\rm dir}.
\end{equation}
Straightforward evaluation of Eq.~\eqref{pertE} then shows that the perturbative
estimate for the binding energy $\bar E_{\rm -,b}$, cf.~Eq.~\eqref{ebinding},
 is negative for $Z=1$ but positive for $Z>1$.  Hence a bound state is predicted
 already by perturbation theory for $Z>1$, in contrast to the case $Z=1$.

Moreover, for $Z>1$, our variational method turns out to be
 restricted to rather small values of $\alpha$. 
 Table \ref{tableZ2singlet} summarizes the rescaled 
binding energies, $\bar E_{\rm -,b}$,  and the optimal charges, $Z_{I,O}$, 
for $Z=2$ and several values of $\alpha$. 
We here consider only the regime $\alpha\le 0.2$ to ensure that the optimal charges remain well below the singular value $Z_{\rm crit}=1/(2\alpha)$. 

For $Z>1$, we also find a bound state in the 
spin-triplet channel.  However,  the variational wave function used here for 
the triplet state is probably not the most appropriate. In the case of the helium atom,
textbooks \cite{bransden}  show that the simplest wave function for the lowest triplet 
state combines the single-particle ground state and the first excited state. 
By analogy, for our 2D case, a better choice for the triplet case might be 
to take instead of Eq.~\eqref{cdans} the ansatz
\begin{equation} 
\Phi_{j=0}= \Psi_{0,\frac{1}{2},I}({\bf r}_1) \Psi_{1,-\frac{1}{2},O}({\bf r}_2) 
-\Psi_{1,-\frac{1}{2},O}({\bf r}_1) \Psi_{0,\frac{1}{2},I}({\bf r}_2), 
\label{J=0}
\end{equation}
which is an eigenstate of the total angular momentum operator $J_z=J_z(1)+J_z(2)$ 
with eigenvalue $j=0$. Another option is 
\begin{equation} 
\Phi_{j=1}= \Psi_{0,\frac{1}{2},I}({\bf r}_1) \Psi_{1,\frac{1}{2},O}({\bf r}_2) -
\Psi_{1,\frac{1}{2},O}({\bf r}_1) \Psi_{0,\frac{1}{2},I}({\bf r}_2).
\label{J=1}
\end{equation}
In the absence of Coulomb interactions, this state has the same energy as
$\Phi_{j=0}$ but Coulomb interactions will mix them. Therefore
a variational approach should take  into account both of them. 
However, this analysis goes beyond the scope of this paper. 

\begin{table} 
    \begin{tabular}{|c|c|c|c|}
    \hline
       $\alpha$	& 	 $\bar E_{\rm -,b}$	& $Z_O$		& $Z_I$ 	\\ \hline\hline
       $0.01$	&	 $7.524$ 		& $1.142$		& $2.266$	\\ \hline
       $0.05$	&	 $7.591$ 		& $1.148$		& $2.261$ \\ \hline
       $0.10$	&	 $7.817$ 		& $1.169$		& $2.243$	 \\  \hline
       $0.15$	&	 $8.265$ 		& $1.210$		& $2.207$	 \\  \hline
       $0.20$	&	 $9.136$ 		& $1.292$		& $2.141$  \\   \hline
      \end{tabular}
 \caption{
 Rescaled binding energy $\bar E_{\rm -,b}$, see Eq.~\eqref{ebinding}, 
 for the two-particle bound state in the spin singlet sector for $Z=2$  and 
several values of $\alpha$.  We also specify the  effective charges $Z_O$ and $Z_I$ 
minimizing the energy functional.}        
\label{tableZ2singlet}
\end{table}
 
\subsection{Topological band gap}
\label{sec4b}

Let us next consider the case of a topological gap,
$H_{\rm so} = \Delta \sigma_zs_z$, see Eq.~\eqref{somass}, 
where we set $\tau_z=1$ as we shall still assume that both quasiparticles are in the same valley. 
(For related studies of the case without Coulomb impurity, see 
Refs.~\cite{rodin,zhong, trushin}.)
Since the total Hamiltonian now does not commute with ${\bf S}^2$ anymore, we must 
distinguish whether the two quasiparticles have same or opposite spin projections, 
$S_z=\pm 1$ or $S_z=0$.
If both quasiparticles have the same spin projection (e.g., $S_z=1$),  we are back to the 
case discussed in Sec.~\ref{sec3} but with $\epsilon=+1$ (spin triplet). 
Within our variational approach for $Z=1$, there is no stable bound state.  

Turning now to $S_z=0$, we cannot separately consider
spin singlet and triplet states. A natural way to construct the Chandrasekhar-Dirac 
variational wave function is as follows.
We consider the 2D subspace spanned by the wave functions
\begin{eqnarray}\nonumber
\Phi_1 & = & \Psi_{I,\uparrow}({\bf r}_1) | \uparrow \rangle_1 \otimes\Psi_{O,\downarrow}({\bf r}_2)  
 | \downarrow \rangle_2 \\
&& - \Psi_{O,\downarrow}({\bf r}_1) |\downarrow \rangle_1 \otimes \Psi_{I,\uparrow}({\bf r}_2) | \uparrow \rangle_2 
\label{chSO},\\
\Phi_2  & =& \Psi_{I,\downarrow}({\bf r}_1) | \downarrow \rangle_1 \otimes
\Psi_{O,\uparrow}({\bf r}_2)  | \uparrow \rangle_2  \nonumber \\
&&-  \Psi_{O,\uparrow}({\bf r}_1) |\uparrow \rangle_1 \otimes \Psi_{I,\downarrow}({\bf r}_2) | \downarrow \rangle_2 ,
\nonumber
\end{eqnarray}
where spin and orbital degrees of freedom are entangled. In Eq.~\eqref{chSO}, $|s\rangle_l$ 
is the eigenstate of $s_z$ with eigenvalue $s=\pm=\uparrow/\downarrow$ for
particle $l=1,2$, and
$\Psi_{\lambda,s}$ refers to the normalized ground state of the 
single-particle Hamiltonian
\begin{equation}
H_{\lambda,s}=-i{\bm \sigma} \cdot \nabla + s \Delta \sigma_z  -\frac{Z_\lambda\alpha}{r}, 
\end{equation}
with  $\lambda=I,O$. Again, 
$Z_{I,O}$ are variational parameters, and the valley part is kept implicit.
Since the valley part is assumed symmetric, we have to choose 
antisymmetric combinations in Eq.~\eqref{chSO}. 

With the ground state of $H_{\lambda,+}$ given by
\begin{equation}
\Psi_{\lambda,\uparrow} \sim r^{\gamma_\lambda-1/2} e^{-p_\lambda r} 
\left( \begin{array}{c}1 \\i \kappa_\lambda e^{i\theta}
\end{array}\right) ,
\end{equation}
we directly obtain 
\begin{equation}
\Psi_{\lambda,\downarrow}=-i
\sigma_y (\Psi_{\lambda,\uparrow})^* \sim 
r^{\gamma_\lambda-1/2} e^{-p_\lambda r} 
\left( \begin{array}{c}i \kappa_\lambda e^{-i\theta} \\
1\end{array}\right) 
\label{antiunit}
\end{equation}
as ground state of $H_{\lambda,-}$ with the same energy,  
where $\langle \Psi_{I,\uparrow} | \Psi_{O,\downarrow}\rangle=0$.
In the subspace spanned by the states $\Phi_{1,2}$, the Hamiltonian
eigenvalue problem thus reduces to the problem of finding solutions of 
the secular equation
\begin{equation}
\det \left(\begin{array}{cc}
H_{11} -2 E & H_{12} - \Sigma_{12} E \\
H_{21} - \Sigma_{21} E  & H_{22}-2 E
\end{array} \right) =0,
\end{equation}
where we use the notation (with $a,b=1,2$)
\begin{equation}
H_{ab} = \langle \Phi_{a} | H | \Phi_b \rangle ,\qquad
\Sigma_{ab}  =  \langle \Phi_a |  \Phi_b \rangle.
\end{equation}
Using the results of Sec.~\ref{sec3} and noting that 
the single-particle matrix elements are independent of the spin projection, 
we find $\Sigma_{11}=\Sigma_{22}=2$ and $\Sigma_{12}=\Sigma_{21}=-2S^2$. The 
roots of the secular equation are given by
\begin{equation} \label{enfunctSO}
E_\pm(Z_I,Z_O)= \frac{ 2H_{11} -H_{12}\Sigma_{12} \pm 
\left|  2 H_{12} - H_{11} \Sigma_{12} \right| }{4-\Sigma_{12}^2},
\end{equation}
It turns out that this expression has the same structure as the energy functional 
in Eq.~(\ref{energyfunct}), and the corresponding
results therefore apply again. We conclude that the topological band gap
caused by Eq.~\eqref{somass} does not imply different bound-state energies as
compared to the topologically trivial band gap resulting from Eq.~\eqref{diracmass}. 

\subsection{Different valleys}
 \label{sec4c}

So far we have assumed that the two quasiparticles occupy the same valley state, 
and we thus only have a trivial double degeneracy of the bound state. We shall
now briefly discuss the case in which the two quasiparticles live in different valleys
(cf.~Ref.~\cite{entin} for a setting without Coulomb impurity).
To properly address this situation,  we first recall that the Dirac-Weyl spinors represent
the slowly varying parts of the electronic wave function. 
The complete wave function is obtained by multiplying
these spinors by the appropriate Bloch wave at the $K$ or $K'$ point.
In the continuum description, one neglects the 
overlap between wave functions in opposite valleys,  
$\langle K | K' \rangle = 0$. (Going beyond this approximation
would require a study of the lattice model.)  
Second, we observe that $H_0$ in Eq.~\eqref{h0dirac} does not
commute with the total squared valley operator 
${\bf T}^2$, where ${\bf T}={\bm \tau}_1 + {\bm \tau}_2$. This fact must be 
taken into account when building the appropriate Chandrasekhar-Dirac
wave function. Finally, the two-body Coulomb interaction potential has the same form 
as for  two quasiparticles with the same valley 
quantum number, see App.~\ref{appB}.
A straightforward calculation as in Sec.~\ref{sec4b} then shows
that the resulting energy functional has the same structure as before. 
Therefore the optimal binding energy coincides with the one for both quasiparticles  
in the same valley.

\section{Observables}
\label{sec5}

In this section, we turn to a discussion of the probability density and 
of the pair distribution function for the bound state. We consider the most interesting case 
$Z=1$ and focus on the Dirac band gap term in Eq.~\eqref{diracmass} 
for the two-body spin singlet state.  However, 
using the results in Sec.~\ref{sec4}, it is straightforward to obtain 
corresponding results also for other cases of interest.

\subsection{Probability density}

We start by calculating the probability density for the two-particle bound state. 
The density operator is 
$\hat \rho({\bf r}) =\sum_{l=1,2} \delta({\bf r} -{\bf r}_l),$
and thus the probability density is given by 
\begin{eqnarray}
\rho({\bf r})&= &\frac{1}{\langle \Phi| \Phi \rangle}
\int d{\bf r}_1 d{\bf r}_2 \Phi^\dagger({\bf r}_1,{\bf r}_2) \, \hat \rho({\bf r}) \,
 \Phi({\bf r}_1,{\bf r}_2)  \nonumber \\
 & =&\int  d{\bf r}_2 
\frac{\left|  \Phi({\bf r},{\bf r}_2)  \right|^2}{1+S^2}  .
\end{eqnarray}
The integral can be evaluated exactly, where the result for $\rho({\bf r})$ 
does not depend  on the polar angle $\theta$. It is then convenient to consider  
the radial density,   
\begin{eqnarray}
P(r) & = & r \int d\theta \rho({\bf r}) = \frac{1}{1+S^2} \Bigl[ 
   \sum_\lambda \frac{(2p_\lambda)^{2\gamma_\lambda+1}}{\Gamma(2\gamma_\lambda+1)}   r^{2\gamma_\lambda}e^{-2p_\lambda r}   \nonumber \\
   &+& 2S^2    \frac{(p_I+p_O)^{\gamma_I+\gamma_O+1}}{\Gamma(\gamma_I+\gamma_O+1)}    r^{\gamma_I+\gamma_O} e^{-(p_I+p_O) r}
   \Bigr],  \label{prr}
\end{eqnarray}
with normalization $\int_0^\infty dr P(r)=2$.
The result is illustrated for $Z=1$ and several values of $\alpha$ in Fig.~\ref{probdensity}.
With respect to the nonrelativistic case, we observe that relativistic
effects tend to enhance the probability at small distance from the 
Coulomb impurity.   Since the radial density can be probed by STM techniques,
the result can be matched to the analytical result in Eq.~\eqref{prr}. Thereby
one can hope to extract, e.g., the value of the fine structure constant $\alpha$.

\begin{figure}[t]
\includegraphics[width=0.5\textwidth]{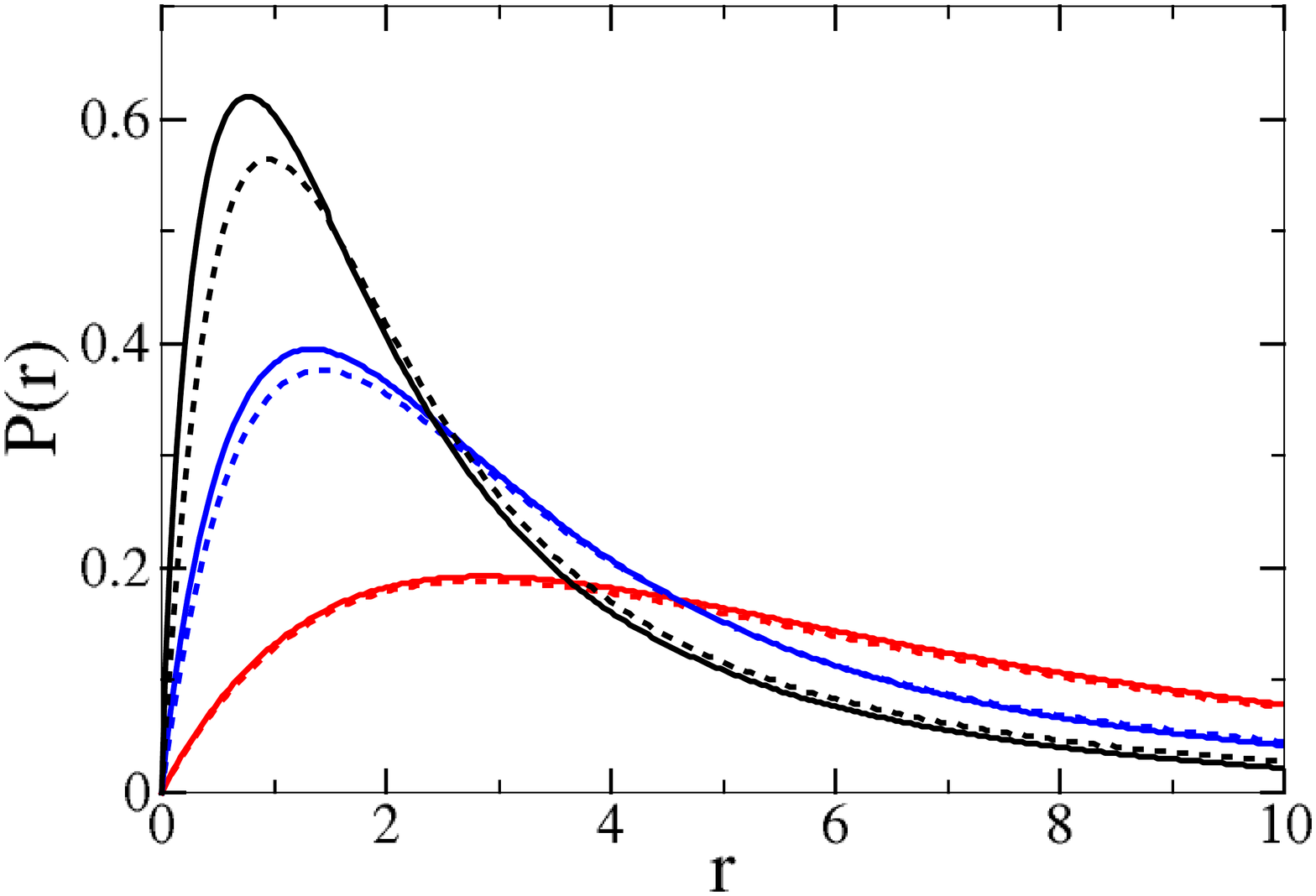}
\caption{Radial density $P(r)$ vs distance $r$ from the impurity (in units of $\hbar v_F/\Delta$) of the spin-singlet two-body bound state, cf.~Eq.~\eqref{prr}, for $Z=1$ and several values of $\alpha$.
Solid lines correspond to the relativistic case for $\alpha=0.1,~0.2,$ and $0.3$,
shown in red, blue, and black color (from bottom to top), respectively. 
Dashed lines indicate the corresponding nonrelativistic results. 
}
\label{probdensity}
\end{figure}

\begin{figure}
\includegraphics[width=0.5\textwidth]{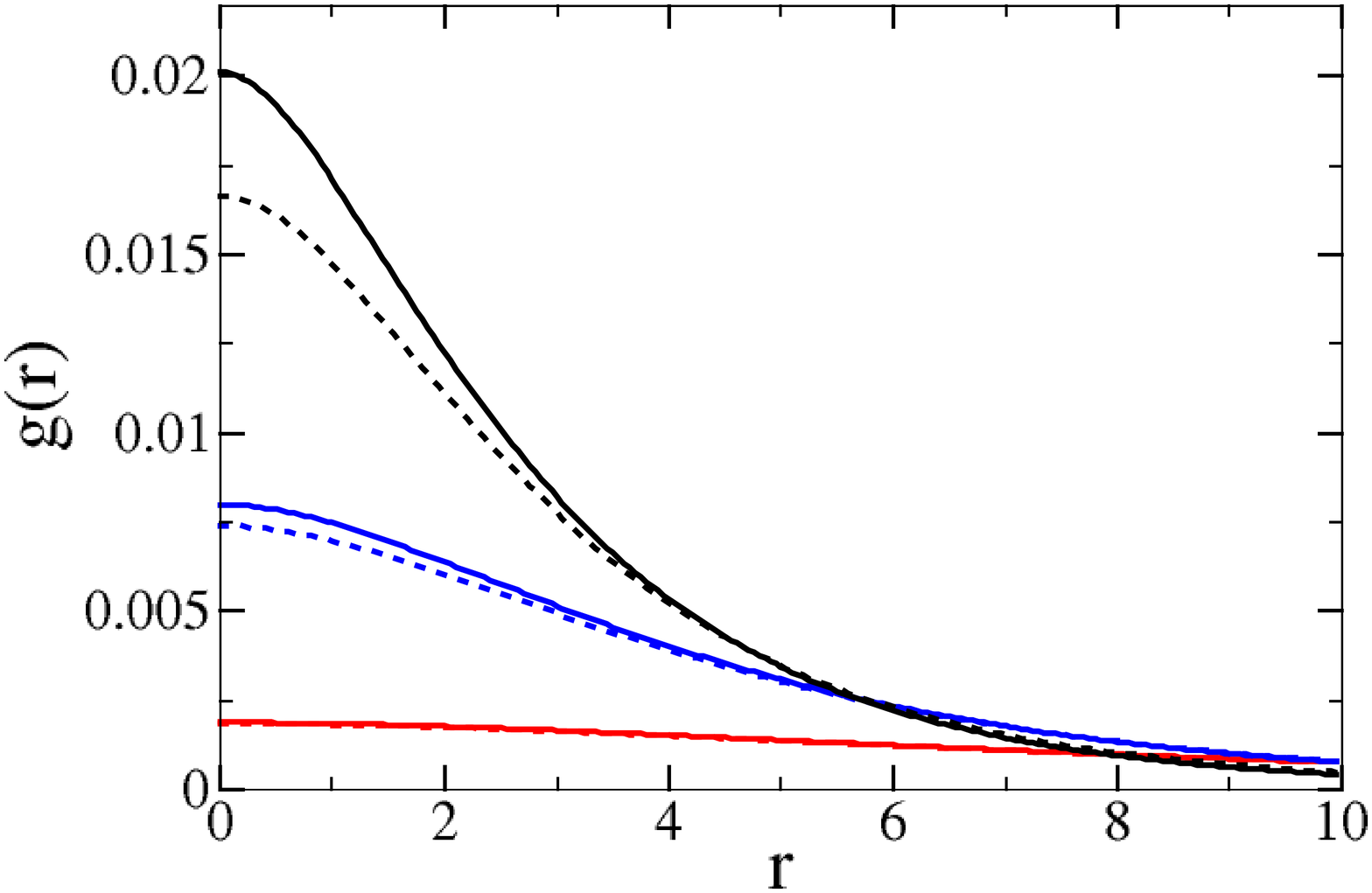}
\caption{Radial profile of the pair distribution function, $g(r)$, vs interparticle distance $r$ (in units of $\hbar v_F/\Delta$) for $Z=1$ and several values of $\alpha$. 
Solid lines correspond to the relativistic case for $\alpha=0.1,~0.2,$ and $0.3$,
shown in red, blue, and black color (from bottom to top), respectively. 
Dashed lines indicate the corresponding nonrelativistic results.}
\label{pairdistr}
\end{figure}

\subsection{Pair distribution function}

Next  we turn to the pair distribution function of the two-particle bound state,
\begin{eqnarray}
g({\bf r}) &=& \frac12\left\langle \sum_{i\neq j } \delta({\bf r}-({\bf r}_i-{\bf r}_j) 
\right\rangle \nonumber \\ &=& \int d{\bf r}_1
 \frac{ | \Phi({\bf r}_1-{\bf r}/2, {\bf r}_1 + {\bf r}/2 ) |^2}{\langle \Phi | \Phi \rangle}. \label{grr}
\end{eqnarray}
The result again turns out to be independent of the polar angle, $g=g(r)$, and is illustrated in Fig.~\ref{pairdistr}.   
The pair distribution function can be experimentally obtained 
by a statistical analysis of STM images, see, e.g., Ref.~\cite{Ertl}, and can provide
additional information about the existence and the properties of the two-body bound
state predicted here.

\section{Concluding remarks}
\label{sec6}

In this work, we have studied the two-particle bound state problem for
gapped graphene in the presence of a Coulomb impurity.
We have shown that a variational approach, using the 
projected Hamiltonian and  Chandrasekhar-Dirac spinors as trial  
wave functions, predicts the existence of at least one bound state. We found that
in contrast to the Schr\"odinger case, the variational energy functional is not
a homogeneous function of  the coupling constant $\alpha$. As a consequence,
the optimal values of the variational parameters $Z_\lambda$ depend on 
$\alpha$, and the optimal binding energy has a more complicated
functional dependence on $\alpha$. In particular,  
the binding energy increases with respect 
to the nonrelativistic case. Moreover, we have determined the 
relativistic corrections to the probability density and to 
the pair probability density.  
The predicted two-body bound state can thereby be accessed experimentally, e.g., 
by means of STM techniques.  

Finally, as a possibility for future theoretical work, it would 
be interesting to diagonalize
 the projected many-body Hamiltonian for graphene in a large basis 
set in order to compute the ground state energy without recourse to 
variational wave functions. This route can also provide information about low-lying 
 excited resonant levels, which in turn are expected to exhibit Fano lineshapes 
 when probed in transport or by STM methods.

\acknowledgments
We thank H.~Siedentop for discussions. This work was supported by the network SPP 1459 (Grant No.~EG 96/8-2) of the Deutsche Forschungsgemeinschaft (Bonn).

\appendix
\section{Relativistic 2D hydrogen atom}
\label{appA}

In order to keep the paper self-contained,  we here
collect known results for the single-particle Dirac-Coulomb problem in graphene,
see, e.g., Refs.~\cite{rmp1,kotov,novikov,gamayun}.
The massive Dirac-Weyl Hamiltonian with a Coulomb impurity of charge
$Ze$ reads ($\hbar=v_F=1$) 
\begin{equation}
H_{\rm D} = -i{\bm \sigma} \cdot \nabla + \Delta  \sigma_z -\frac{Z\alpha}{r}.
\label{diracweyl}
\end{equation}
The Hamiltonian \eqref{diracweyl}  is exactly solvable, and the bound-state orbitals can be found, e.g., in Refs.~\cite{novikov, gamayun}. 
Following the notation of Ref.~\cite{novikov}, in polar coordinates ($r,\theta$) they are given by
\begin{equation}
 \Psi_{n,j}(r,\theta) = {\cal N}_{n,j} \rho^{\gamma-\frac{1}{2}} e^{-\frac{\rho}{2}}
\left( \begin{array}{r}
 (\varphi_1+ c \varphi_2) e^{i(j-\frac{1}{2})\theta}\\
i\kappa (\varphi_1 - c \varphi_2)  e^{i(j+\frac{1}{2})\theta} 
 \end{array} \right) .
\label{boundS}
 \end{equation}
The half-integer index $j$ denotes
 the eigenvalue of the total angular momentum operator
$J_z=-i\partial_\theta +\sigma_z/2$.
The integer index $n\ge 0$  is the principal quantum number, where the  
energy eigenvalues are given by
 \begin{equation} \label{energieeigen}
E_{n,j} = \frac{\Delta}{\sqrt{1+ \frac{Z^2\alpha^2}{(n+\gamma)^2}}}.
\end{equation}
We use the notation
 \begin{eqnarray}
 \rho &= &2 p r, \qquad
 p=\sqrt{\Delta^2-E^2},\\ \nonumber
 \gamma&=&\sqrt{j^2-Z^2\alpha^2},  \quad
 \kappa = \sqrt{\frac{\Delta-E}{\Delta+E}},\\
  \nonumber c &= &\frac{\gamma -\frac{Z\alpha E}{p}}{j+\frac{Z\alpha\Delta}{p}}= 
 \frac{j - \frac{Z\alpha  \Delta}{p}}{\gamma + \frac{Z\alpha E}{p}}.
 \end{eqnarray}
 The functions $\varphi_{1,2}$ in Eq.~\eqref{boundS} 
 are confluent hypergeometric functions of the first kind \cite{olver},
\begin{eqnarray}
\varphi_1(\rho)& = &M(\gamma -Z\alpha E/p,2\gamma+1,\rho),\\ \nonumber
\varphi_2(\rho) &= &M( \gamma +1-Z\alpha E/p,2\gamma+1,\rho).
\end{eqnarray}
Finally, ${\cal N}_{n,j}$ is a normalization constant such that
$\int rdrd\theta | \Psi_{n,j} |^2 =1$. Explicitly, one finds
\begin{equation}
{\cal N}_{n,j}= \frac{(-1)^n p^\frac{3}{2}}{\Delta \Gamma(2\gamma+1)}
\sqrt{\frac{\Gamma(2\gamma+1+n)(\Delta+E)(j+\frac{Z\alpha\Delta}{p})}
{2\pi Z\alpha \, n!}}.
\end{equation}

The energy eigenvalues  \eqref{energieeigen} now follow from the condition
$\gamma-Z\alpha E/p = - n$, such that the
wave functions are normalizable. The confluent hypergeometric functions then reduce
to generalized Laguerre polynomials \cite{olver},
\begin{equation}
M(-n,2\gamma+1,\rho) = \frac{\Gamma(n+1) \Gamma(2\gamma+1)}{\Gamma(2\gamma+1+n)}L^{2\gamma}_n(\rho).
\end{equation}
The $n>0$ bound states are doubly degenerate, $E_{n,j}=E_{n,-j}$, while the
$n=0$ bound states exist only for $j>0$. Technically,
this is due to the fact that for $n=0$, i.e.,
$\gamma-Z\alpha E/p=0$, $\varphi_2$ grows exponentially $\sim e^{\rho}$, 
and the corresponding solution is 
admissible only for $c=0$. This in turn occurs only for $j>0$, while $c=-1$ for $j<0$.

The lowest-energy bound state is given by 
\begin{equation}
\Psi_{0,\frac{1}{2}} = {\cal N}_{0,\frac{1}{2}} \rho^{\gamma-\frac{1}{2}} e^{-\frac{\rho}{2}}\left(\begin{array}{l}1\\
i\kappa e^{i\theta}\end{array}\right),
\end{equation}
with the energy
\begin{equation}
E_{0,\frac{1}{2}}=2\Delta\gamma=\Delta \sqrt{1-4Z^2 \alpha^2},
\label{lowesteigenvalue}
\end{equation}
and the  normalization factor
\begin{equation}
{\cal N}_{0,\frac{1}{2}} = 
2Z\alpha\Delta \sqrt{ \frac{2\gamma+1}{\pi \Gamma(2\gamma+1) }}.
\end{equation}

States in the continuum spectrum, $|E| > \Delta$,
can be obtained by means of analytic continuation in $E$, see
Ref.~\cite{novikov}. One finds that the states are given 
by Eq.~(\ref{boundS}) with the  substitutions
\begin{eqnarray}
p & =& \sqrt{\Delta^2-E^2}  \rightarrow -i\sqrt{E^2-\Delta^2} \equiv -i \tilde p, \\
\rho & = &2pr \rightarrow  -2 i \tilde p r, \nonumber \\ \nonumber
c &=&\frac{\gamma -Z\alpha E/p}{j +Z \alpha\Delta/p} \to
  \frac{\gamma -iZ\alpha E/\tilde p}{j 
+ iZ\alpha\Delta/\tilde p} \equiv e^{-2i\xi} \equiv\tilde c\\ \nonumber
\kappa & =& \sqrt{\frac{\Delta - E}{\Delta+E}} \rightarrow -i\, \text{sgn}(E) 
\,\sqrt{\frac{E-\Delta}{E+\Delta}} \\ \nonumber 
& \equiv & - i  \, \tilde \kappa  = -i \frac{\tilde p}{E+\Delta} 
= -i \frac{E-\Delta}{\tilde p}.
\end{eqnarray}
Explicitly, we get
\begin{equation}
\Psi_{E,j}= {\cal N}_{E,j} r^{\gamma-1/2} e^{ i \tilde p r}
\left( \begin{array}{r}
 (\varphi_1+ \tilde c \varphi_2) e^{i(j-\frac{1}{2})\theta}\\
  \tilde \kappa (\varphi_1 - \tilde c \varphi_2)  e^{i(j+\frac{1}{2})\theta} 
 \end{array}\right) ,
\label{continuumS}
\end{equation}
where 
\begin{eqnarray}
\varphi_1 &=& M(\gamma - iZ\alpha E/\tilde p,2\gamma +1,-2i\tilde p r), \\
\varphi_2 &=& M(\gamma+1 - iZ\alpha E/\tilde p,2\gamma +1,-2i\tilde p r).\nonumber
\end{eqnarray}
The normalization factor $ {\cal N}_{E,j}$
follows by matching the asymptotic behavior of Eq.~\eqref{continuumS} to  
that of free  spherical spinors \cite{novikov}, and reads
 \begin{equation}
{\cal N}_{E,j} = \left( \frac{|E+\Delta|}{2|E|}  \right)^{\frac{1}{2}} 
\frac{|\Gamma(1+\gamma + \frac{iZ\alpha E}{\tilde p})|}
{\sqrt{2\pi}\Gamma(2\gamma+1)}
e^{\frac{\pi Z\alpha E}{2\tilde p}} (2\tilde p)^\gamma  e^{i\xi}.
\label{continuumN}
 \end{equation}
The spinors then satisfy the identity
\begin{equation}
 \int d{\bf r}\, \Psi^\dagger_{E,j}({\bf r}) \Psi_{E',j'}({\bf r}) =  
 2\pi \delta(\tilde p-\tilde p') \Theta(EE')\delta_{jj'} ,
 \end{equation}
where $\Theta$ is the Heaviside step function.
Therefore the resolution of the identity for the (projected) 
Coulomb-Dirac problem reads
\begin{eqnarray}
{\mathbb 1} &=& \sum_{n,j} |\Psi_{n,j}\rangle \langle \Psi_{n,j}| +\\ \nonumber
&+& \sum_j \int_{|E|>\Delta} \frac{dE}{2\pi\sqrt{E^2-\Delta^2}}
|\Psi_{E,j}\rangle \langle \Psi_{E,j}|.
\end{eqnarray}

Finally, we provide the overlaps between our variational wave function 
$|\Psi_\lambda\rangle$ in Eq.~\eqref{eigenspin} and the
bound states as well as with continuum states. Using the identity 
\begin{equation}
\int_0^\infty e^{-\lambda r} r^\nu M(a,c,k r) dr = 
\frac{\Gamma(\nu+1)}{\lambda^{\nu+1}} {}_2F_1(a,\nu+1,c,k/\lambda),
\end{equation}
where ${}_2F_1(a,b,c,z)$ is the hypergeometric function \cite{olver},
 the overlap integrals with bound states are given by
\begin{widetext}
\begin{eqnarray}
C_{n,j}&=&
\langle \Psi_\lambda | \Psi_{n,j} \rangle = 2\pi \delta_{j,\frac{1}{2}} 
{\cal N}_\lambda {\cal N}_{n,j} \frac{
\Gamma(\gamma_\lambda+\gamma+1)}
{(p_\lambda+p)^{\gamma_\lambda+\gamma+1}}
\Biggl[ \left( 1+  \frac{2\alpha_\lambda (\Delta-E_{n,j})}{(2\gamma_\lambda+1)p} 
\right) {}_2F_1\left(-n,\gamma_\lambda +\gamma+1, 2\gamma+1, 
\frac{2p}{p_\lambda+p}\right)  \nonumber \\ 
&- &  \frac{n}{j+ \alpha\Delta/p }
\left( 1-  \frac{2\alpha_\lambda (\Delta-E_{n,j})}{(2\gamma_\lambda+1)p} \right) 
{}_2F_1\left(-n+1,\gamma_\lambda +\gamma+1, 2\gamma+1, \frac{2p}{p_\lambda+p}\right) \Biggr],
\label{overlapBS}
\end{eqnarray}
while the overlap with continuum states is given by
\begin{eqnarray}
C_j(E)&=&\langle \Psi_\lambda  |\Psi_{E,j} \rangle  =
2\pi \delta_{j,\frac{1}{2}} {\cal N}_\lambda {\cal N}_{E,j} \int_0^\infty dr \, r^{\gamma_\lambda+\gamma}
e^{-(2\alpha_\lambda- i \tilde p)r} \left[(1- i \tilde \kappa \kappa_\lambda) \varphi_1 +c (1 +i \tilde \kappa\kappa_\lambda)\varphi_2
\right],\nonumber \\
&=& 2\pi \delta_{j,\frac{1}{2}} {\cal N}_\lambda {\cal N}_{E,\frac{1}{2}}  
\frac{\Gamma(\gamma_\lambda+\gamma+1)}{(p_\lambda-i\tilde p)^{\gamma_\lambda+\gamma+1}}
\Biggl[ \left(1- i\frac{2\alpha_\lambda(E-\Delta)}{(2\gamma_\lambda+1)\tilde p} \right) 
{}_2F_1\left(\gamma- i\alpha E/\tilde p,\gamma_\lambda+\gamma+1,2\gamma+1, 
\frac{-2i\tilde p}{p_\lambda - i \tilde p}\right)  \nonumber \\
&+&   \frac{\gamma- i\alpha E/ \tilde p}{j+i \Delta\alpha/\tilde p}
\left(1+ i \frac{2\alpha_\lambda(E-\Delta)}{(2\gamma_\lambda+1)\tilde p} \right) 
{}_2F_1\left(\gamma- i\alpha E/\tilde p+1,\gamma_\lambda+\gamma+1,2\gamma+1, 
\frac{-2 i\tilde p}{p_\lambda-i\tilde p}\right)\Biggr].
\label{overlapCS}
\end{eqnarray}
\end{widetext}
Using the identity ${}_2F_1(a,b,b,z)=(1-z)^{-a}$,
one can check that for $Z_\lambda=Z$, the overlaps 
$C_{n,j}$ reduce to $\delta_{n,0}\delta_{j,1/2}$
and that the overlaps $C_j(E)$ vanish. Furthermore,
since the $|\Psi_\lambda\rangle$ states are normalized to unity, the expansion
coefficients satisfy the identity
\begin{equation}
\sum_{n,j} |C_{n,j}|^2 + \sum_j 
\int \frac{dE}{2\pi \sqrt{E^2-\Delta^2}} |C_{j}(E)|^2 =1.
\end{equation}

\section{Coulomb interaction} 
\label{appB}

Here we briefly discuss the form of the two-body interaction used in our analysis.  
In general, the many-body Coulomb interaction is given by 
\begin{equation}
H_{\rm int} =\frac{1}{2}\int d{\bf r}_1 d{\bf r}_2 \Psi_{s_1}^\dagger({\bf r}_1) 
\Psi^\dagger_{s_2}({\bf r}_2) V_{\rm 2b}({\bf r}_{12})
\Psi_{s_2}({\bf r}_2) \Psi_{s_1}({\bf r}_1) 
\end{equation}
where $\Psi_s(\bf r)$ is the field operator with spin projection $s=\pm 1$, and
the sum over spin projections is understood.
For graphene, the field operator in the continuum limit 
can be decomposed  into valley components,
\[
\Psi_s({\bf r})  \simeq \sum_{\tau=\pm} e^{i\tau {\bf K}\cdot {\bf r}}\Psi_{s,\tau}({\bf r}),
\]
where $\pm {\bf K}$ are the two independent Fermi momenta (Dirac points) 
and $\tau=\pm$ is the valley index. 

Correspondingly, $H_{\rm int}$ decomposes into several terms,
$H_{\rm int} \simeq \sum_{j=1}^5H^{(j)}_{\rm int}$,
where (the sum over repeated spin and valley 
indices is understood, and $\bar \tau= -\tau$)
\begin{widetext}
\begin{eqnarray}
H^{(1)}_{\rm int}&=& \frac{1}{2} \int d{\bf r}_1 d{\bf r}_2 \; \Psi_{s_1,\tau}^\dagger({\bf r}_1) 
\Psi^\dagger_{s_2,\tau}({\bf r}_2) V_{\rm 2b}({\bf r}_{12})
\Psi_{s_2,\tau}({\bf r}_2) \Psi_{s_1,\tau}({\bf r}_1), \nonumber\\
H^{(2)}_{\rm int}&=& \frac{1}{2} \int d{\bf r}_1 d{\bf r}_2 \; \Psi_{s_1,\tau}^\dagger({\bf r}_1) 
\Psi^\dagger_{s_2,\bar \tau}({\bf r}_2) V_{\rm 2b}({\bf r}_{12})
\Psi_{s_2, \bar \tau}({\bf r}_2) \Psi_{s_1,\tau}({\bf r}_1), \nonumber\\
H^{(3)}_{\rm int} &= &\frac{\tilde V_{\rm 2b}(2{\bf K})}{2} \int d{\bf R} \;
\Psi_{s_1,\tau}^\dagger({\bf R})
\Psi^\dagger_{s_2, \bar \tau}({\bf R})
\Psi_{s_2, \tau}({\bf R}) \Psi_{s_1, \bar \tau} ({\bf R}), \label{H3} \\
H^{(4)}_{\rm int}& =& \frac{1}{2} \int d{\bf r}_1 d{\bf r}_2 \; \Psi_{s_1,\tau}^\dagger({\bf r}_1) \Psi^\dagger_{s_2, \tau}({\bf r}_2)
e^{-i2\tau {\bf K}\cdot ( {\bf {\bf r}_1 + {\bf r}_2})} V_{\rm 2b}({\bf r}_{12})
\Psi_{s_2, \bar \tau}({\bf r}_2) \Psi_{s_1, \bar \tau} ({\bf r}_1),  \nonumber \\
H^{(5)}_{\rm int}& =& \frac{\tilde V_{\rm 2b}({\bf K})}{2} \int d{\bf R} \; 
\Psi_{s_1,\tau_1}^\dagger({\bf R})  
\Psi^\dagger_{s_2, \tau_2}({\bf R})
e^{-i2\tau_1 {\bf K}\cdot{\bf R}}
\Psi_{s_2, \tau_2}({\bf R}) \Psi_{s_1, \bar \tau_1} ({\bf R}) \nonumber \\ 
&+&\frac{\tilde V_{\rm 2b}({\bf K})}{2} \int d{\bf R} \;
\Psi_{s_1,\tau_1}^\dagger({\bf R})  
\Psi^\dagger_{s_2, \tau_2}({\bf R})
e^{-i2\tau_2 {\bf K}\cdot{\bf R}} 
\Psi_{s_2, \bar \tau_2}({\bf R}) \Psi_{s_1,  \tau_1} ({\bf R}) . \label{H5}
\end{eqnarray}
\end{widetext}
In order to obtain Eqs.~\eqref{H3} and \eqref{H5}, we have 
switched to center-of-mass and relative coordinates, 
${\bf R}=({\bf r}_1 + {\bf r}_2 )/2$ and 
${\bf r}={\bf r}_1-{\bf r}_2$, respectively, and subsequently  
integrated over the relative coordinate.  Furthermore,
$\tilde V_{\rm 2b}({\bf q})$ denotes 
the Fourier component of the Coulomb potential, where ${\bf q}={\bf K}$ 
or ${\bf q}=2{\bf K}$.

For our problem, we expect that the dominant matrix elements of $H_{\rm int}$ 
in the two-particle subspace are those due to $H^{(1)}_{\rm int}$ if both particles 
belong to the same valley, or those due to $H^{(2)}_{\rm int}$ if they belong to 
opposite valleys. The matrix elements of all other terms are suppressed
either by a small coupling constant $[H^{(3)}_{\rm int}]$, 
 due to rapidly fluctuating exponential factors in the integral
$[H^{(4)}_{\rm int}]$, or by both mechanisms together $[H^{(5)}_{\rm int}]$.

\end{document}